\documentclass[a4paper,11pt]{article}
\usepackage{pos}
\usepackage{graphicx}

\title{Studies of New Physics in $B^0_q-\bar B^0_q$ Mixing and Implications for Leptonic Decays}

\author[a,b]{Kristof De Bruyn}
\author[a,c]{Robert Fleischer}
\author*[a,d]{Eleftheria Malami}
\author[e]{Philine van Vliet}

\affiliation[a]{Nikhef,\\ 
Science Park 105, 1098 XG Amsterdam, Netherlands}

\affiliation[b]{Van Swinderen Institute for Particle Physics and Gravity, University of Groningen,\\
9747 Groningen, Netherlands}

\affiliation[c]{Faculty of Science, Vrije Universiteit Amsterdam,\\
1081 HV Amsterdam, Netherlands}

\affiliation[d]{ Center for Particle Physics Siegen (CPPS), Theoretische Physik 1,
Universität Siegen,\\
D-57068 Siegen, Germany}

\affiliation[e]{Deutsches Elektronen-Synchrotron DESY,\\
Notkestr. 85, 22607 Hamburg, Germany}

\emailAdd{Eleftheria.Malami@uni-siegen.de}

\abstract{The phenomenon of $B^0_q$-$\bar{B}^0_q$ mixing ($q=d,s$) provides a sensitive probe for physics beyond the Standard Model. We have a careful look at the determination of the Unitarity Triangle apex, which is needed for the Standard Model predictions of the $B_q$ mixing parameters, and explore how much space for New Physics is left through the current data. We study the impact of tensions between inclusive and exclusive determinations of the CKM matrix elements $|V_{ub}|$ and $|V_{cb}|$, and  focus on the $\gamma$ angle extraction. We present various future scenarios and discuss the application of these results for leptonic rare $B$ decays, which allows us to minimise the CKM parameter impact in the New Physics searches. Performing future projections, we explore and illustrate the impact of increased precision on key input quantities. It will be exciting to see how more precise data in the future high-precision era of flavour physics can lead to a much sharper picture.}

\FullConference{%
  8th Symposium on Prospects in the Physics of Discrete Symmetries (DISCRETE 2022)\\
  7-11 November, 2022\\
  Baden-Baden, Germany
}

\begin{document}
\maketitle

\section{Introduction}
The phenomenon of $B^0_q$-$\bar{B}^0_q$ mixing (where $q=d,s$) arises only from loop processes in the Standard Model (SM) and is sensitive to possible New Physics (NP) contributions, which could enter the loop topologies or even at the tree level, for instance in $Z'$ models. Associated to the mixing phenomenon are the mixing parameters and the CP-violating phases for which we have impressive experimental data. In this presentation, we follow Ref.~\cite{DeBruyn:2022zhw} and explore the space allowed for NP by current measurements and the state-of-the-art parameters. In addition, we point out interesting connections to the studies of leptonic rare $B$ decays.

In order to determine the parameter space of possible NP effects to $B_q^0$--$\bar{B}_q^0$ mixing, we have to compare the SM predictions of the mixing parameters with the corresponding experimental values. For these SM predictions, a careful analysis of the Unitarity Triangle (UT) apex is required. We pay special attention to the different determinations of the Cabibbo-Kobayashi-Maskawa (CKM) parameters and the tensions that arise between the extractions of the $|V_{ub}|$ and $|V_{cb}|$ matrix elements through inclusive and exclusive semileptonic $B$ meson decays. These longstanding tensions have a profound impact on the whole analysis. 

\section{Unitarity Triangle}
Using the parametrisation of the Particle Data Group (PDG), the UT apex is given as \cite{ParticleDataGroup:2022pth}:
\begin{equation}
R_b \ e^{i \gamma}  = \bar{\rho} + i \bar{\eta}\:, \qquad 
    \bar\rho \equiv \left[1-({\lambda^2}/{2})\right] \rho\:,\qquad
    \bar\eta \equiv \left[1-({\lambda^2}/{2})\right] \eta\:.
\end{equation}
Here, $\rho$, $\eta$ and $\lambda$ are the Wolfenstein parameters \cite{Wolfenstein:1983yz,Buras:1994ec}, $R_b$ is the side from the origin to the apex of the UT, defined with the help of the CKM matrix elements $\lambda \equiv |V_{us}|, |V_{ub}|$ and $|V_{cb}|$ as:
\begin{equation}\label{eq:Rb}
    R_b \equiv \left(1-\frac{\lambda^2}{2}\right)\frac{1}{\lambda}\left|\frac{V_{ub}}{V_{cb}}\right|
    = \sqrt{\bar\rho\,^2 + \bar\eta\,^2}\:,
\end{equation}
 and $\gamma \equiv \arg\left(-{V_{ud}^{\phantom{*}}V_{ub}^*}/{V_{cd}^{\phantom{*}}V_{cb}^*}\right)\:$ is the angle between the $R_b$ side and the UT basis.

\subsection{Determining the UT Apex Utilising $\gamma$ and $R_b$}
In this subsection, we work in the SM and are interested in obtaining the UT apex in a way that is not affected by  possible NP in $B^0_q$-$\bar{B}^0_q$ mixing. One way of determining the apex is utilising the side $R_b$ and the angle $\gamma$, which can both be determined from decays that proceed only via tree decays. The value of $\gamma$ can be determined either from $B \to DK$ decays or from a $B \to \pi\pi,\ \rho\pi,\ \rho\rho$ isospin analysis. 

More specifically, one option is to use the time-dependent $B^0_s \to D_s^{\mp}K^{\pm}$ system, where mixing-induced CP violation plays a key role. Through interference effects caused by $B^0_q$-$\bar{B}^0_q$ mixing, the CP asymmetry parameters allow the determination of $\phi_s + \gamma$, where $\phi_s$ is the $B^0_s$-$\bar{B}^0_s$ mixing phase. Since $\phi_s$ is determined through the $B^0_s \to J/\psi \phi$ channel, including penguin corrections \cite{Barel:2020jvf,Barel:2022wfr}, $\gamma$ can be obtained in a theoretically clean way \cite{Fleischer:2021cct,Fleischer:2021cwb}. However, the surprisingly large value arising in this case still needs to be further explored.  An alternative way of getting the $\gamma$ value is using the time-independent $B \to DK$ transitions, where the sensitivity to $\gamma$ comes from direct CP violation \cite{LHCb:2021dcr}. Last but not least, another interesting system is provided by $B \to \pi\pi,\ \rho\pi,\ \rho\rho$ modes \cite{Gronau:1990ka,Charles:2017evz}, which usually are used to determine $\alpha$ from an isospin analysis. Actually this value corresponds to $\gamma$ when we use the $B^0_d$-$\bar{B}^0_d$ mixing phase $\phi_d$, determined from $B^0_d \to J/\psi K^0$ \cite{Barel:2020jvf,Barel:2022wfr}, taking penguin effects into account. Thus, we can convert the result $\phi_d + 2\gamma$ into $\gamma$. The value from the latter case is in good agreement with the one coming from $B \to DK$ modes. Therefore, for our analysis, we average these two results \cite{DeBruyn:2022zhw}:
\begin{equation}\label{eq:gamma_Avg}
    \gamma_{\text{avg}} = (68.4 \pm 3.4)^{\circ}.
\end{equation}

Regarding $R_b$ there are tensions between the various theoretical and experimental approaches. Even though there are different determinations of the $|V_{us}|$ element and the tensions between them are intriguing, they only have a negligible impact on NP studies in neutral $B_q$ mixing. Thus, we choose to work with the value $|V_{us}|= 0.22309 \pm 0.00056$  \cite{Seng:2021nar,Seng:2022wcw}. Contrary to the $|V_{us}|$ case, the deviations between determinations of $ |V_{ub}|$ and $|V_{cb}|$ from inclusive and exclusive semileptonic $B$ decays, which are given as follows \cite{HFLAV:2022pwe,Bordone:2021oof}:
\begin{align}
    |V_{ub}|_{\text{incl}} &= (4.19 \pm 0.17) \times 10^{-3}\:, &\ &  |V_{ub}|_{\text{excl}} = (3.51 \pm 0.12) \times 10^{-3}\:,&\ & {\text{differing by }}3.9\ \sigma,\\
    |V_{cb}|_{\text{incl}} &= (42.16 \pm 0.50) \times 10^{-3}\:, &\ &  |V_{cb}|_{\text{excl}} = (39.10 \pm 0.50) \times 10^{-3}\:,  &\ &  {\text{differing by }} 4.3\ \sigma,
\end{align}
have a significant impact on the allowed parameter space for NP in $B^0_q$-$\bar{B}^0_q$ mixing. Trying to understand and resolve these tensions, another case is studied in the literature \cite{Ricciardi:2021shl,Bordone:2021oof,Bordone:2019guc,Buras:2022wpw}, which is a hybrid scenario combining the exclusive $|V_{ub}|$ with the inclusive $|V_{cb}|$ determination. Therefore, we consider for the rest of our analysis all these three cases. The corresponding $R_b$ results are:
\begin{equation}\label{eq:Rb_val}
    R_{b,\text{incl}} = 0.434 \pm 0.018\:, \qquad
    R_{b,\text{excl}} = 0.392 \pm 0.014\:, \qquad
    R_{b,\text{hybrid}} = 0.364 \pm 0.013\:.
\end{equation}

\begin{figure}[t!]
    \centering
    \includegraphics[width=0.41\textwidth]{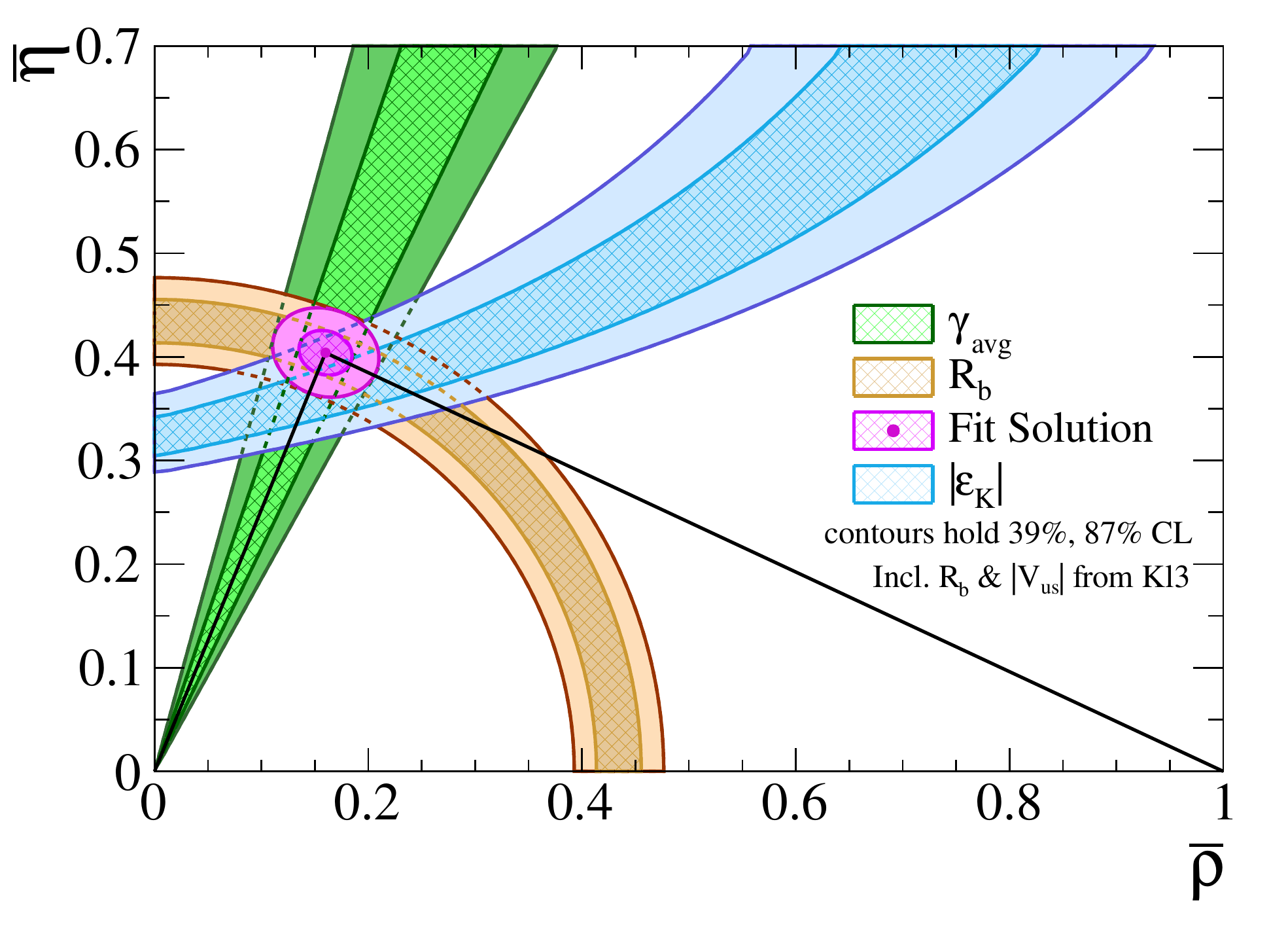}
    \includegraphics[width=0.41\textwidth]{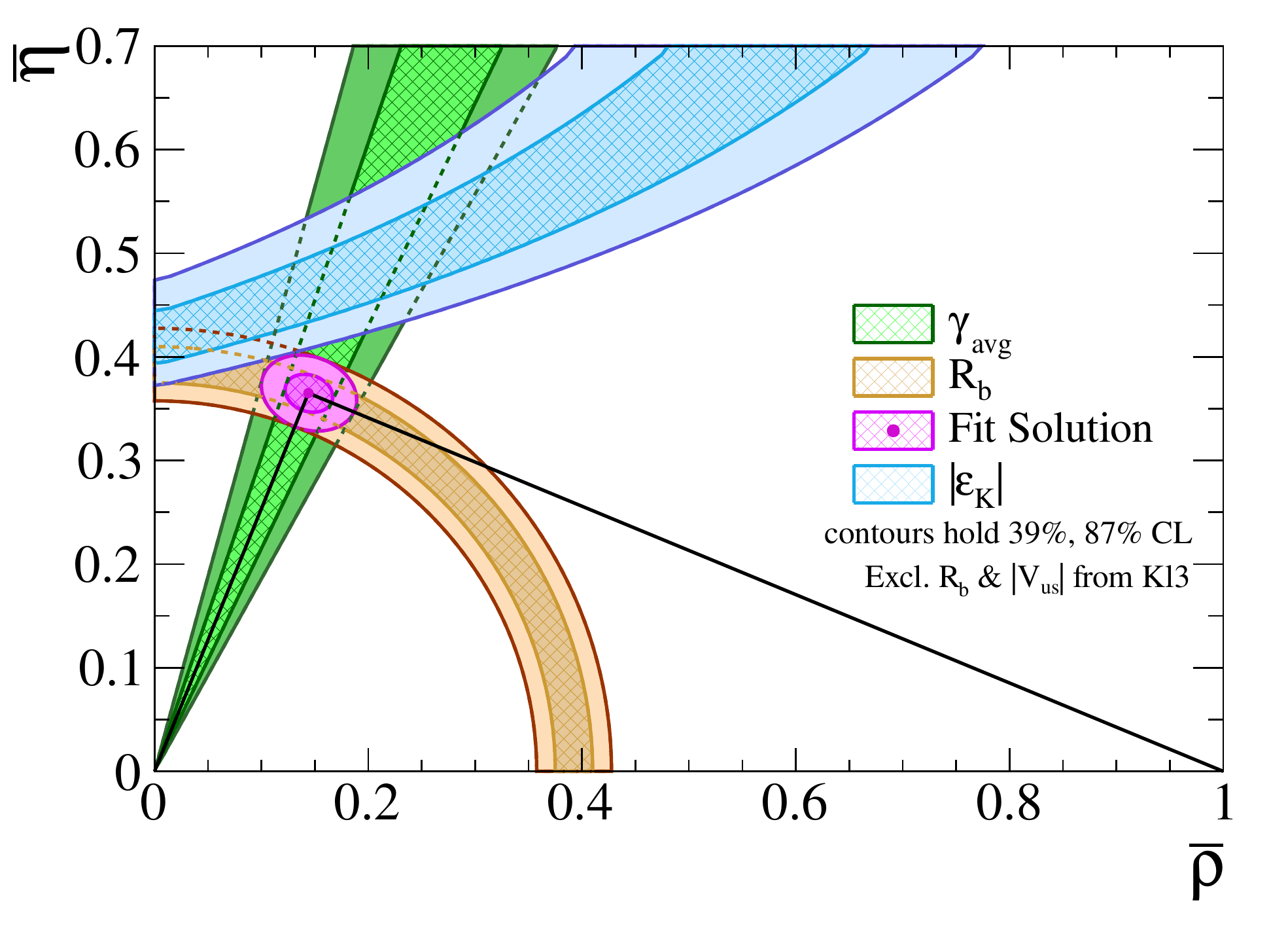}
    \includegraphics[width=0.41\textwidth]{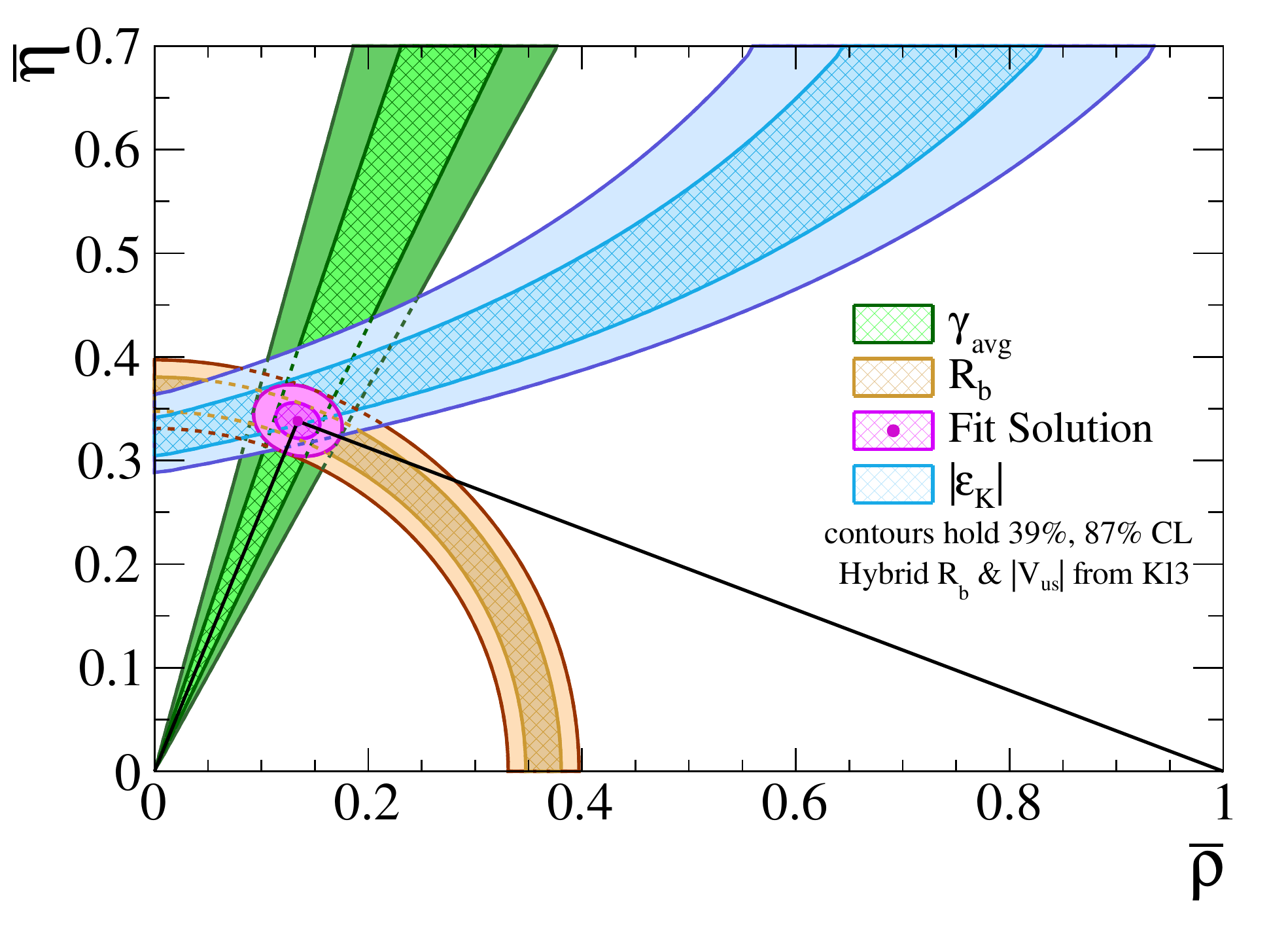}
    \caption{Determination of the UT apex from the $R_b$ and $\gamma$ measurements for the inclusive (left), exclusive (right) and hybrid (botttom) case \cite{DeBruyn:2022zhw}.}
    \label{fig:UT_apex}
\end{figure}

Making a fit to $R_b$ and $\gamma$, the UT apex is determined \cite{DeBruyn:2022zhw}: 
\begin{align}
    \text{Incl.} & &
    \bar\rho & = 0.160 \pm 0.025 \:, & 
    \bar\eta & = 0.404 \pm 0.022\:, \label{eq:UT_apex_I3}\\
    \text{Excl.}  & &
    \bar\rho & = 0.144 \pm 0.022 \:, & 
    \bar\eta & = 0.365 \pm 0.018\:, \label{eq:UT_apex_E3}\\
    \text{Hybrid } & &
    \bar\rho & = 0.134 \pm 0.021 \:, & 
    \bar\eta & = 0.338 \pm 0.017\:.\label{eq:UT_apex_H3}
\end{align}
The results are illustrated in Fig.~\ref{fig:UT_apex}. The plot also shows the hyperbola coming from the $|\varepsilon_K|$ observable, which is related to indirect CP violation in the neutral kaon system and is highly sensitive to the $|V_{cb}|$ numerical value. The hybrid case gives the most consistent picture of the UT apex within the SM, which illustrates the strong dependence on $|V_{cb}|$. In the future, this could help us to understand the inclusive-exclusive puzzle, if NP in the kaon system can be controlled or ignored.

\subsection{Determining the UT Apex Utilising $R_b$ and $R_t$}
An alternative way of determining the UT apex is utilising the $R_t$ side, which is defined as:
\begin{equation}\label{eq:Rt}
    R_t \equiv \left| {V_{td}V_{tb}}/{V_{cd}V_{cb}} \right| 
    = \sqrt{(1 - \bar\rho)^2 + \bar\eta\,^2} .
\end{equation}
In this case, only information on the two UT sides $R_b$ and $R_t$ is required without needing any information from $\gamma$. However, in order to get the $R_t$, we have to assume SM expressions for the mixing parameters $\Delta m_{d}$ and $\Delta m_{s}$. The numerical predictions are given in \cite{DeBruyn:2022zhw}.

The side $R_t $ can be written as
\begin{equation}\label{eq:Rt}
    R_t    = \frac{1}{\lambda}\left| \frac{V_{td}}{V_{ts}} \right| \left[1-\frac{\lambda^2}{2}\left(1-2\bar\rho\right)\right] + \mathcal{O}\left(\lambda^4\right)\:, 
    \end{equation}
where
\begin{equation}
 \left| \frac{V_{td}}{V_{ts}} \right| = \xi \sqrt{\frac{m_{B_s} \Delta m_{d}^{\text{SM}}}{m_{B_d} \Delta m_{s}^{\text{SM}}}}.
\end{equation}
Here the SU(3)-breaking parameter $\xi$ is the ratio of bag parameters and decay constants of the $B_d$ and the $B_s$ systems that can be calculated on the lattice. The advantage of the ratio is that uncertainties cancel, making it cleaner than using individual results. 

Making a fit to the $R_b$ and $R_t$ sides, we obtain \cite{DeBruyn:2022zhw}:
\begin{align}
    \text{Incl.} & &
    \bar\rho & = 0.180 \pm 0.014 \:, & 
    \bar\eta & = 0.395 \pm 0.020\:, \label{eq:Rt_apex_I3} \\
    \text{Excl.} & &
    \bar\rho & = 0.163 \pm 0.013 \:, & 
    \bar\eta & = 0.357 \pm 0.017\:, \label{eq:Rt_apex_E3} \\
    \text{Hybrid }& &
    \bar\rho & = 0.153 \pm 0.013 \:, & 
    \bar\eta & = 0.330 \pm 0.016\:. \label{eq:Rt_apex_H3}
\end{align}
We note that the UT apex determinations relying on $\gamma$ are a factor $2$ less precise than those without information from $\gamma$. However, the determination through $R_b$ and $R_t$ requires the SM expressions of  $\Delta m_{d}$ and $\Delta m_{s}$, thus ignores possible NP contributions in $B^0_q$-$\bar{B}^0_q$ mixing.

\section{NP in $B^0_q$-$\bar{B}^0_q$ mixing}
The neutral $B_q$-meson mixing is a sensitive phenomenon for NP. In order to quantify its impact, we introduce NP parameters $\kappa_q$, which describes the size of the NP effects, and $\sigma_q$, which is a complex phase accounting for additional CP-violating effects. The generalised expressions of the mixing parameters take the following form \cite{Ball:2006xx}:
\begin{align}
    \Delta m_{q} & = \Delta m_{q}^{\text{SM}} \left|1 + \kappa_{q} e^{i\sigma_{q}}\right|\:, \label{eq:NP_mix_Dmq}\\
    \phi_{q} & = \phi_{q}^{\text{SM}} + \phi_{q}^{\text{NP}} = \phi_{q}^{\text{SM}}  + \arg\left(1 + \kappa_{q} e^{i\sigma_{q}}\right)\:. \label{eq:NP_mix_phiq}
\end{align}
This is a model independent parametrization. Utilising these relations, we explore two different NP scenarios; the first one is the most general case and the second one assumes Flavour Universal NP (FUNP) \cite{DeBruyn:2022zhw}.
\begin{figure}[t!]
    \centering
    \includegraphics[width=0.41\textwidth]{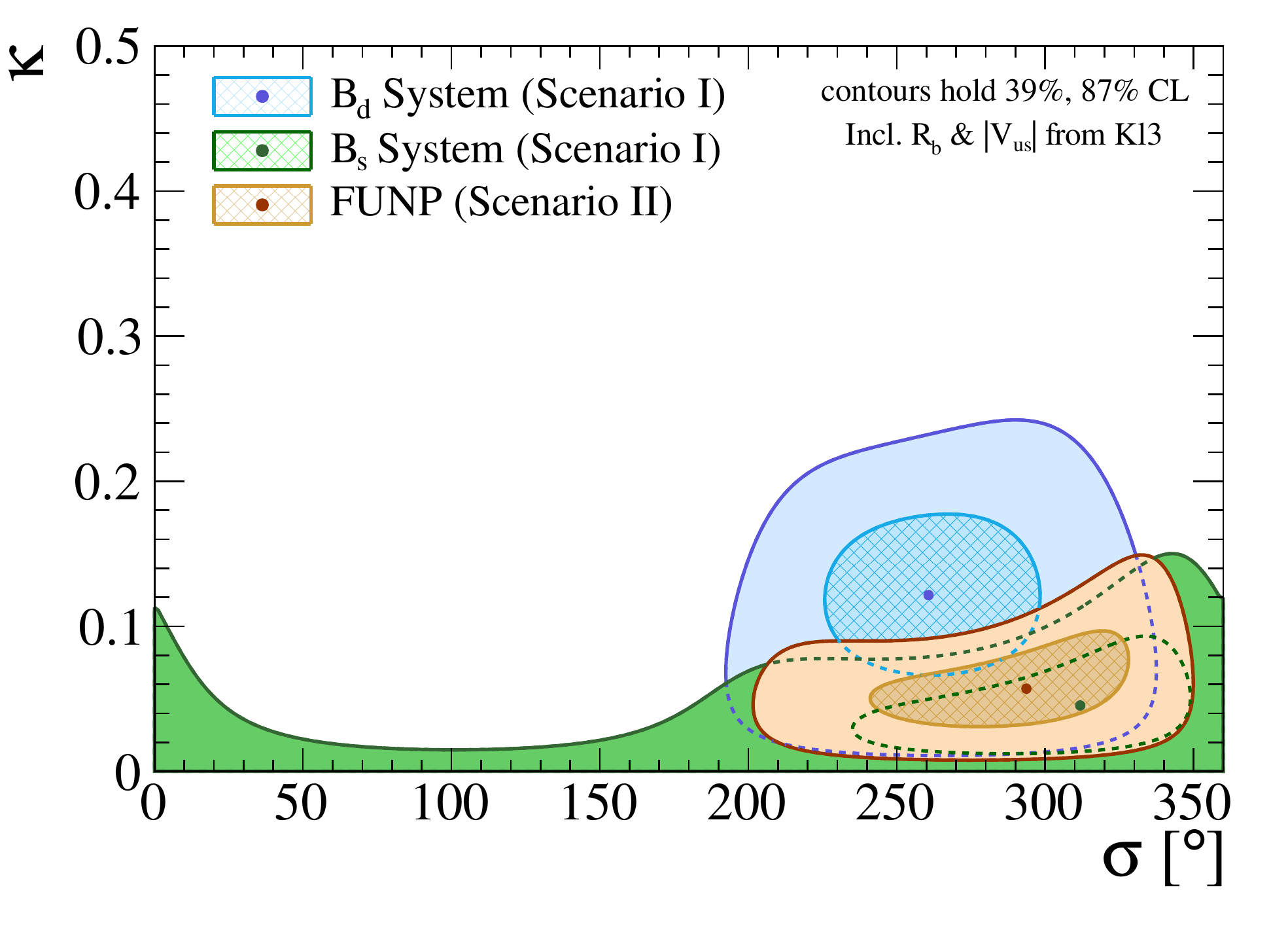}
    \includegraphics[width=0.41\textwidth]{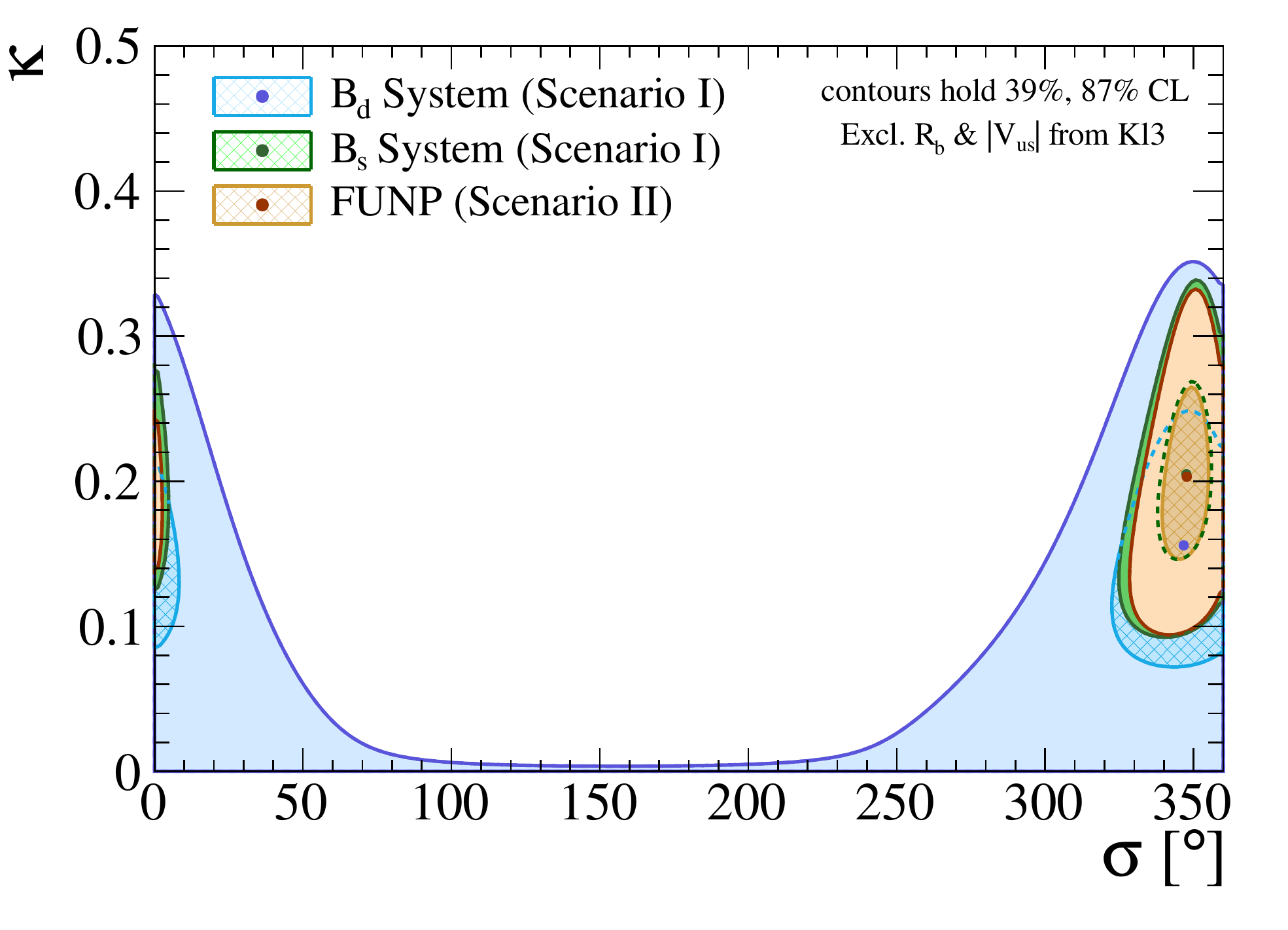}    
    \includegraphics[width=0.41\textwidth]{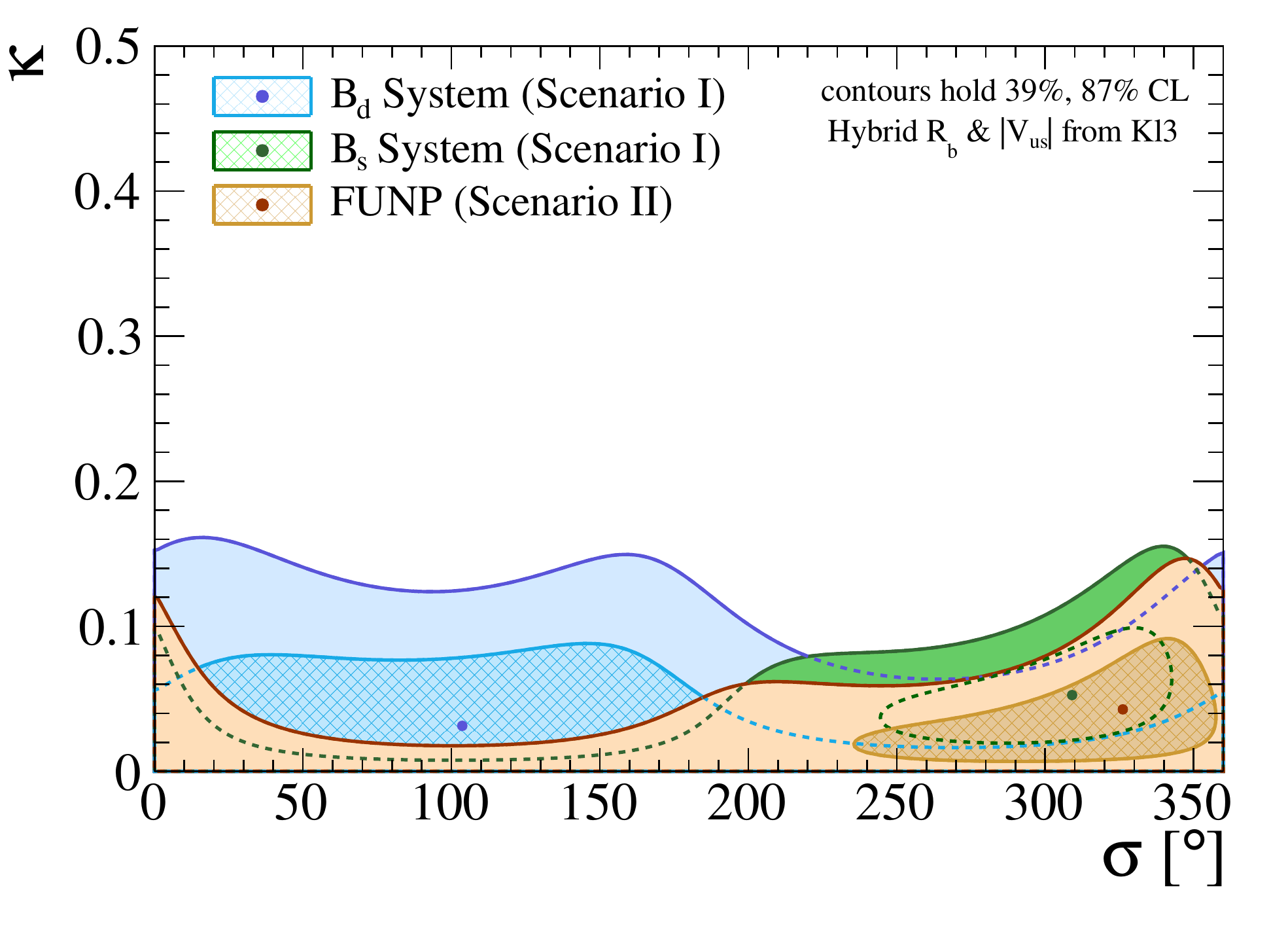}
    \caption{Comparing Scenario I and Scenario II fits for $\kappa_q$ and $\sigma_q$ for the inclusive (left), exclusive (right) and hybrid (bottom) case \cite{DeBruyn:2022zhw}.}
    \label{fig:NPS1vsS2}
\end{figure}

Let us firstly discuss the general case, namely Scenario I. The only assumption here is that there is no NP in the angle $\gamma$ and $R_b$. The determination from $R_b$ and $\gamma$ does not rely on information from mixing. We make use of this determination to obtain the UT apex, which we then need for getting the SM predictions for the mixing parameters $ \Delta m_{q}$ and $ \phi_{q}$.  Comparing them with their measured values, we can constrain the NP parameters. Here, the NP parameters $(\kappa_d, \sigma_d)$ and $(\kappa_s, \sigma_s)$ are determined independently from each other.

In the second case, Scenario II,  we have the FUNP assumption where we consider that the NP contributions are equal in the $B_d$ and $B_s$ systems, thus $(\kappa_d, \sigma_d) = (\kappa_s, \sigma_s)$. This is not a Minimal Flavour Violation scenario but it can be realised in NP models with $U(2)$ symmetry \cite{Barbieri:2012uh,Charles:2013aka}. The UT apex fit relies on $R_b$ and $R_t$, without using $\gamma$ information, therefore possible NP in the angle $\gamma$ will not affect the findings. Comparing the two scenarios, we have a test of the FUNP assumption and we see the impact of the assumptions on the constraints on the parameter space of NP in mixing. Fig.~\ref{fig:NPS1vsS2} illustrates this comparison of the two fits for $\kappa_q$ and $\sigma_q$ for the inclusive, the exclusive and the hybrid cases.

\section{Rare Leptonic Decays  $B_q^0\to\mu^+\mu^-$}
The tensions between the CKM matrix elements have an impact not only on the UT apex determination and possible NP in $B^0_q$-$\bar{B}^0_q$ mixing but also on the branching ratios of rare decays. A key example is the leptonic $B_q^0\to\mu^+\mu^-$ transition. These modes are pure loop processes and helicity suppressed in the SM. This helicity suppression could be lifted by new scalar and pseudoscalar conttributions, therefore putting these decays in an outstanding position to probe NP in this sector. 
As these are decays of neutral $B$ mesons, $B^0_q$-$\bar{B}^0_q$ mixing enters and leads to subtleties concerning the measurement of the experimental branching ratio and comparison with the theoretical prediction \cite{DeBruyn:2012wk}. However, NP in $B^0_s$-$\bar{B}^0_s$ mixing is included through the experimental values of the mixing parameters.

The SM predictions require information on $|V_{ts}|$ which we determine through $|V_{cb}|$, which again depends on inclusive and exclusive determinations. In order to minimise the dependence on $|V_{cb}|$ and the UT apex, we create the following ratio with the $B_s$ mass difference $\Delta m_s$  \cite{Buras:2003td,Buras:2021nns,Bobeth:2021cxm}:
\begin{equation}
    \mathcal{R}_{s\mu} \equiv {\bar{\mathcal{B}}(B_s\to\mu^+\mu^-)}/{\Delta m_s} \:.
\end{equation}
Using this ratio, we can eliminate the leading dependence on the CKM elements but we have to correct for the possible NP contributions to $B^0_q$-$\bar{B}^0_q$ mixing. This is now possible following our analysis in \cite{DeBruyn:2022zhw}.

So, we include NP effects in $\Delta m_s$ and then we can use the ratio $\mathcal{R}_{s\mu}$ to constrain NP in the scalar and pseudoscalar sector. We obtain the generalised expression:
\begin{equation}\label{eq:Rsmu}
    \mathcal{R}_{s\mu} =
    \mathcal{R}_{s\mu}^{\text{SM}} \times
    \frac{1 + \mathcal{A}^{\mu\mu}_{\Delta \Gamma_s} y_s}{1 + y_s}
    \frac{|P_{\mu\mu}^{s}|^2 + |S_{\mu\mu}^{s}|^2}{\sqrt{1 + 2 \kappa_s \cos\sigma_s + \kappa_s^2}}\:,
\end{equation}
with $P_{\mu\mu}^{s} \equiv |P_{\mu\mu}^{s}|e^{i\varphi_P}$, $S_{\mu\mu}^{s} \equiv |S_{\mu\mu}^{s}|e^{i\varphi_S}$, where $\varphi_P$, $\varphi_S$ are CP-violating phases, and the observable $\mathcal{A}^{\mu\mu}_{\Delta \Gamma_s}$ in terms of the NP phase $ \phi_s^{\text{NP}}$: \begin{equation}\label{eq:Amumu}
    \mathcal{A}_{\Delta \Gamma}^{\mu\mu} = 
    \frac{|P_{\mu\mu}^s|^2\cos(2\varphi_P - \phi_s^{\text{NP}}) - |S_{\mu\mu}^s|^2\cos(2\varphi_S - \phi_s^{\text{NP}})}{|P_{\mu\mu}^s|^2 + |S_{\mu\mu}^s|^2}\:.
\end{equation} The $ \mathcal{R}_{s\mu} $ has only a dependence on the CKM matrix elements through the NP parameters $\kappa_q$ and $\sigma_q$, determined as described above. Therefore, we have another constraint on the scalar and pseudoscalar contributions. The same strategy can be applied to the $B_d^0\to\mu^+\mu^-$ channel once in the future accurate measurements of the branching ratio will become available.

\section{Future Prospects and Final Remarks}
It will be important in the future to achieve improved precision on the NP parameters $\kappa_q$ and $\sigma_q$. In order to get a feeling of the prospects, we assume a hypothetical reduction of $50\%$ on each one of the three input parameters, which are the $|V_{cb}|$, the lattice calculations and the UT apex \cite{DeBruyn:2022zhw}. We obtain interesting findings, which of course depend on these assumptions. In our studies, we demonstrate that in the $B_d$-system the apex plays a limiting factor and in order to fully explore the potentials of this system, progress on the UT apex has to be made. On the other hand, in the $B_s$-system we do not have this situation as the SM prediction of $\phi_s$ is more robust. Therefore, searches of NP in $B^0_s$-$\bar{B}^0_s$ mixing are more promising than in the $B_d$-system but it is of key importance to constrain NP in both systems as much as possible. 

Another essential future prospect is related to the angle $\gamma$. Improved precision on the input measurements might lead to significant discrepancies between the different $\gamma$ determinations due to NP effects. In this case, averaging over the different results, as we did in this analysis, would no longer be justified. Therefore, the UT should then be revisited. Independent information from additional observables would be necessary to resolve such a situation. Exciting new opportunities might come up to search for NP, both in $\gamma$ and in $B^0_q$-$\bar{B}^0_q$ mixing, which is strongly correlated with the UT apex coordinates.

Last but not least, the branching ratios of the $B_q^0\to\mu^+\mu^-$ decays might offer interesting opportunities. The ratio of the branching fractions between $B_d^0\to\mu^+\mu^-$ and $B_s^0\to\mu^+\mu^-$ can provide an alternative way to determine the UT side $R_t$. Another useful application for the ratio of the branching fractions between these channels is the quantity \cite{Fleischer:2017ltw}:
\begin{align}
    U_{\mu\mu}^{ds} 
   \propto \left[ 
    \left|\frac{V_{ts}}{V_{td}}\right|^2
    \frac{\bar{\mathcal{B}}(B_d\to\mu^+\mu^-)}{\bar{\mathcal{B}}(B_s\to\mu^+\mu^-)}
    \right]^{1/2}\:
\end{align}
which requires knowledge of $R_t$ and offers a very powerful test of the SM, where  $U_{\mu \mu}^{ds} = 1$.

In the future, $B^0_q$-$\bar{B}^0_q$ mixing will remain a key element for constraining NP. It will be exciting to see how more precise data in the high-precision era of flavour physics ahead of us can lead to a much sharper picture.

\section*{Acknowledgements}
\noindent
We would like to thank the DISCRETE 2022 organisers for the invitation and for giving us the opportunity to present our studies.
This research has been supported by the Netherlands Organisation for Scientific Research (NWO). PvV acknowledges support from the DFG through the Emmy Noether research project 400570283, and through the German-Israeli Project Cooperation (DIP).

\end{document}